\documentclass[prb,twocolumn,showpacs,superscriptaddress]{revtex4}
\usepackage{graphicx,dcolumn,amsmath}

\begin{document}

\title { Core-level spectroscopy of Si/SiO$_2$ quantum wells: evidence for
confined states }

\author{\firstname{Pierre} \surname{Carrier}} \email[Author to whom
correspondence should be addressed: Email address:\ ]
{Pierre.Carrier@UMontreal.CA} \affiliation{D\'{e}partement de Physique et
Groupe de Recherche en Physique \\et Technologie des Couches Minces (GCM),
Universit\'{e} de Montr\'{e}al, \\ Case Postale 6128,
Succursale~Centre-Ville, Montr\'{e}al, Qu\'{e}bec, Canada H3C 3J7}

\author{\firstname{Z. H.} \surname{Lu}} \affiliation{Department of
Materials Science and Engineering, University of Toronto, Toronto, Canada
M5S 3E4}

\author{\firstname{M. W. C.} \surname{Dharma-wardana}}
\affiliation{Institute for Microstructural Sciences, National Research
Council, Ottawa, Canada K1A 0R6}

\author{\firstname{Laurent J.} \surname{Lewis}}
\affiliation{D\'{e}partement de Physique et Groupe de Recherche en
Physique \\et Technologie des Couches Minces (GCM), Universit\'{e} de
Montr\'{e}al, \\ Case Postale 6128, Succursale~Centre-Ville, Montr\'{e}al,
Qu\'{e}bec, Canada H3C 3J7}

\date{\today}

\begin{abstract}

We present an experimental and theoretical study of the conduction states
of crystalline Si films confined within amorphous SiO$_2$ barriers, using
the Si-$2p$ core-level excitations. The spectral peaks near the conduction
band minimum are compared with the bulk silicon spectrum. In the Si
quantum wells, it is found that the conduction band minimum and the
low-lying peaks undergo a blue shift while all higher peaks \emph{remain
unshifted}. The experimental results are supported by calculations using
recent first-principles structural models for Si/SiO$_2$ superlattices.  
The experimental results suggest that all conduction states up to a given
conduction band offset become confined and blue-shifted while those at
higher energies are not confined and undergo no shift. These results
provide unambiguous evidence that the visible-light emitting properties of
Si/SiO$_2$ structures depend strongly on electron confinement effects.

\end{abstract}


\maketitle

The discovery of luminescence in porous silicon set off strong efforts
towards the fabrication of Si-based light-emitting nanostructures.  Thus,
Si-nanocrystallites, wires, layers, as well as Si/Ge superlattices have
been studied in this context.\cite{mrs97} A study of luminescence in
Si/SiO$_2$ quantum wells (QWs)  was reported by Lu \emph{et al}., where
evidence for quantum confinement was presented.\cite{llb} Many studies of
Si-luminescence have led to similar or quite different conclusions
involving explanations in terms of surface defect centers, self-trapped
excitons, siloxene derivatives and so forth.\cite{kanemistu,mrs95} On the
theory side, while several studies have established the importance of
electron confinement in Si by the SiO$_2$,\cite{pcarDBM,pcarMRS,
garoufalis,jap,pcarFRM} others have emphasized the role of interface
states\cite{Kageshima} and some even questioned the relevance of
confinement in these structures containing amorphous barriers. Also,
experimental studies which attempted to probe the conduction band density
of states (DOS) failed to see clear evidence of electron confinement in
Si/SiO$_2$ QW structures. Meanwhile, the local-energy gap on the SiO$_2$
side has been probed using electron energy-loss spectroscopy (EELS) by
Neaton \emph{et al}.\cite{neaton}

We report in this Letter experimental Si-$2p$ core-level spectroscopy
results coupled with first-principles calculations of
atomistically-detailed Si/SiO$_2$ QW structures to demonstrate electron
confinement and the resulting blue shifting. Confined states are directly
relate to the width of the Si-layer, as opposed for instance to impurity
states that are uncorrelated to any geometric parameter. In Si/SiO$_2$
QWs, only states found within the energy range of an effective
conduction-band offset (CBO) are expected to be confined, while band
states above should remain unconfined.  This scenario is analysed from
both experiment and theory sides.
 \begin{figure}[b]
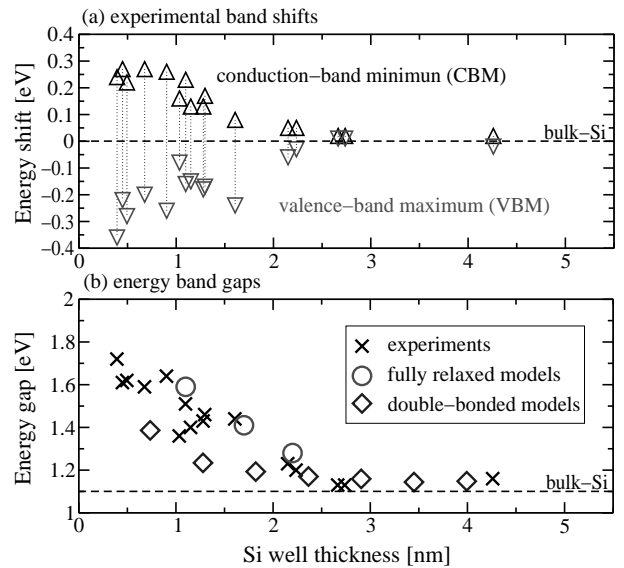

 \includegraphics*[width=8cm]{cbmvbm.eps}
 \includegraphics*[width=8cm]{egap.eps}
 \caption
 { (a) Energy shifts of the conduction-band minimum ($\bigtriangleup$) and
the valence-band maximum ($\bigtriangledown$) of SiO$_2$/Si/SiO$_2$ QW as
a function of the well width.  The dotted lines connect identical samples.  
(b) Experimental energy gaps ({\large\bf $\times$}) as compared with early
theoretical results obtained from idealized confining wells having Si=O
double bonds at interfaces\cite{pcarDBM} ({\Large $\diamond$}) and from
the fully-relaxed models\cite{pcarFRM} ({\Large $\circ$}). }
 \label{xanedat}
 \end{figure}

Very recently, the fabrication of \emph{crystalline} SiO$_2$/Si/SiO$_2$
QWs have been reported by Lu and Grozea.\cite{lg} The QWs having various
thicknesses were examined using a synchrotron source; X-ray photoelectron
spectra (XPS) as well as X-ray absorption near-edge spectra (XANES) were
recorded.  The valence-band maxima (VBM) and conduction-band minima (CBM)
were deduced respectively from the XPS and the XANES.  All data were
calibrated according to an assumed null VBM for crystalline silicon
(c-Si), corresponding to the limit of an infinitely thick silicon well.
The results for the energy shifts of the VBM and the CBM as a function of
the Si well width are summarized in Fig.~\ref{xanedat}(a). The energy gaps
were obtained by summation of the reference energy gap of c-Si (1.12 eV)
with the relative CBMs and VBMs.  The resulting energy gaps are compiled
in Fig.~\ref{xanedat}(b).  The bandgaps are compared with two theoretical
models obtained within the density-functional theory (DFT) and the local
density approximation (LDA), where all bandgaps are raised by 0.6 eV
(corresponding to the difference between the LDA and experimental bandgaps
of c-Si). The theoretical energy gaps for the QWs are found to be in good
agreement with the experiments.
 \begin{figure}[b]
 \includegraphics*[width=8cm]{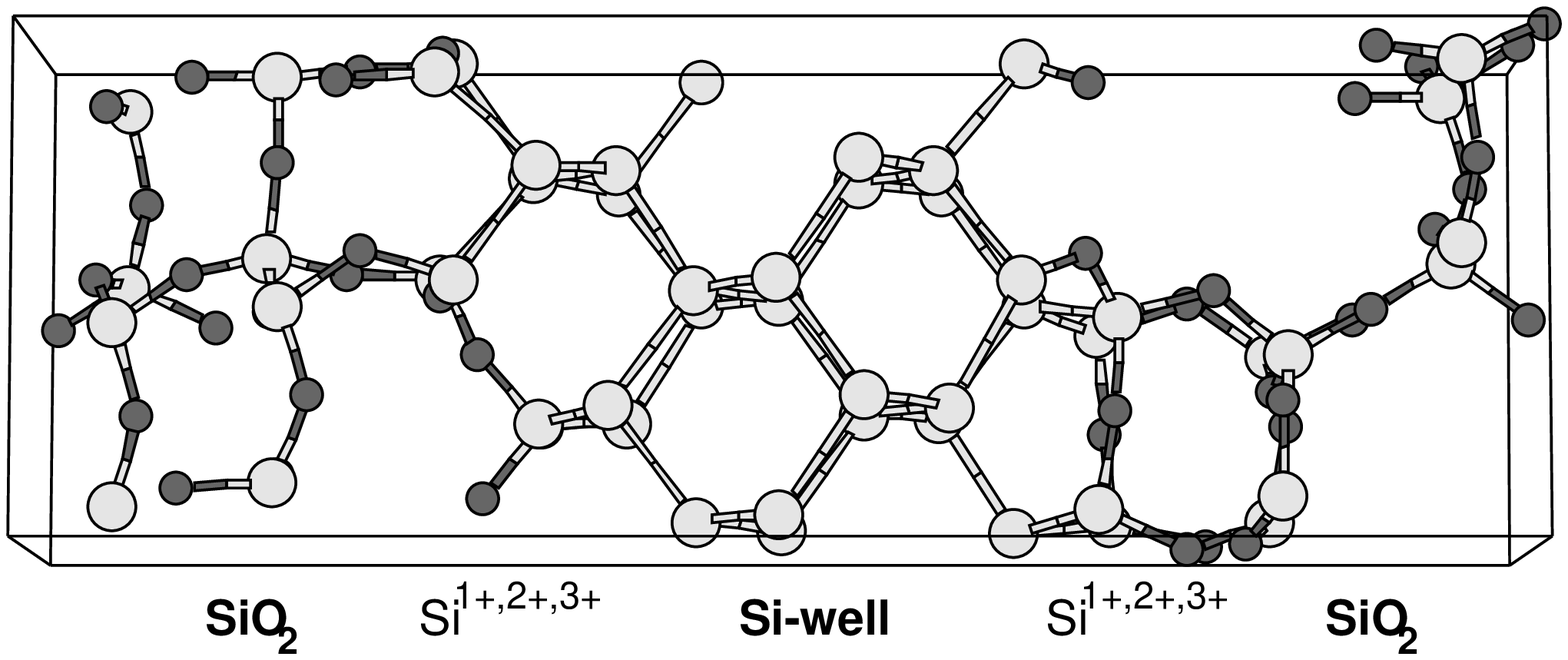}
 \includegraphics*[width=8cm]{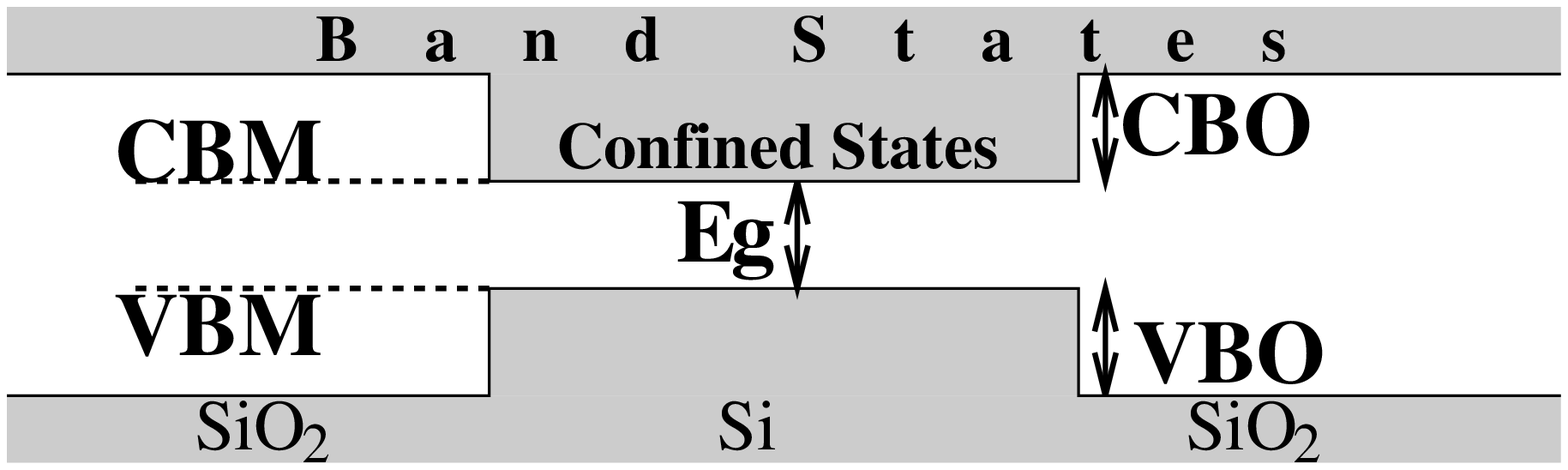}
 \caption
 { The basic unit of the fully-relaxed model (FRM) of the
SiO$_2$/Si/SiO$_2$ structure used in these calculations. Longer units are
obtained by inserting additional Si unit cells in the well region.
Bulk-like Si and SiO$_2$, as well as suboxides Si atomic regions are
referred.  The models provide a theoretical underpinning to the core-level
experiments.  Definitions of the various quantities discussed in the text
are also schematized. The Si-well domain of the model on top and the
electronic ``confined states'' sketched at the bottom are directly
related.}
 \label{FRM}
 \end{figure}

We have carried out atomistically-detailed calculations using a
first-principles-based energy optimized Si/SiO$_2$ interface model, where
Si-well layers are coupled to ``amorphized'' SiO$_2$ layers. The structure
is based on one of the Si/SiO$_2$/{\it vacuum} interface models of
Pasquarello et al,\cite{alfredo1} extended to a SiO$_2$/Si/SiO$_2$ system
by Tit and Dharma-wardana,\cite{jap} and finally fully optimized by
Carrier, Lewis, and Dharma-wardana,\cite{pcarFRM} wherein can be found
calculations of the electronic states, optical matrix elements and
densities of states. The detailed atomic structure of the 1.1 nm QWs is
shown in Fig.~\ref{FRM}.  Definitions of the VBM, CBM and the
corresponding energy gap ($E_g$) evaluated in Fig.~\ref{xanedat}, as well
as definitions of the CBO and VBO, are also schematized.  The three
suboxide Si atomic species (Si$^{1+}$, Si$^{2+}$, and Si$^{3+}$) are
present at the interfaces, as indicated in Fig.~\ref{FRM}.  Complete
details about structural deformations from the bulk can be found in
Ref.~\onlinecite{pcarFRM}.

Consider now a piece of matter which is in an initial state with total
energy $E_i$ together with a probe photon of energy $h\nu_i$. The
interaction leaves the system in a final state of energy $E_f$ together
with an emitted photon of energy $h\nu_f$. The process involves the
absorption of a photon of energy $h\omega$ = $h(\nu_f-\nu_i)$ if no free
electrons are emitted in the final state. By sweeping the probe photon
energy so that transitions from the $2p$ core-states of Si are excited to
the empty states in the conduction band of Si, the energies of the
electronic states in the conduction band can be determined. The energy
change $h\omega=E_f-E_i$ includes all initial-state as well as final-state
interactions and can be calculated using DFT, since the latter provides an
approach to accurate total energies. On the other hand, within the Fermi
Golden rule, the absorption spectrum of the probe photon is given in terms
of matrix elements $|M_{2p,final}|^2$ and joint densities of states (jDOS)
$J(\omega)$ obtained by a convolution of the initial and final densities
of states,
 $
 J(\omega)= \int d\nu D_i(\nu)D_f(\nu+\omega).
 $
 In our case the initial-DOS $D_i$ is that of the Si-$2p$ core-state at
the energy $\epsilon_{2p}$. This is a discrete relativistic doublet of
states, and hence a sum of $\delta$-functions. Thus, the jDOS is
essentially proportional to the sum of pure conduction electron densities
of states. The matrix element connecting the initial and final states will
select only the final states which have $s$- and
$d$-symmetries.\cite{fuggle} The transition rate $T$ is thus determined,
first, by:
 \begin{eqnarray*}
 T(\omega) & = &
  |M_{2p,s}(\omega)|^2 D_s(\epsilon_{2p}+\omega)  \\
       & & \hspace{7pt} + \hspace{3pt}
           |M_{2p,d}(\omega)|^2D_d(\epsilon_{2p}+\omega).
 \end{eqnarray*}
 Second, the electron binding energies of the $2j_{1/2}$ and $2j_{3/2}$
core-states are separated by 0.6 eV (98.8 - 98.2 eV).\cite{fuggle} The
final expression for the Si-$2p$ core-level absorption is thus obtained by
summation of shifted transitions rate $T$:
 \begin{equation}
 A_{2p} = \frac{2}{6}T( \epsilon_{2p}+\omega) + \frac{4}{6}
          T( \epsilon_{2p}+ \mbox{0.6 eV} +\omega). \label{A2p}
 \end{equation}

 The $s$- and $d$-DOS calculated in a straightforward way using Kohn-Sham
methods will have several shortcomings: (a) The absolute position of the
DOS peaks would not be correct due to the well known underevaluation of
the bandgap; (b) the effect of final-state interactions would not be
contained in the standard calculation of the DOS; (c) the type of photon
renormalization effects arising from time-dependent response is not
contained in the standard DOS evaluation.\cite{zangwill} Some of these
shortcomings can be overcome by using fairly simple modifications to the
DOS evaluation, e.g., the Slater-transition-states (STS)
method.\cite{tian, DreizlerGross}
 \begin{figure}
 \includegraphics*[width=8cm]{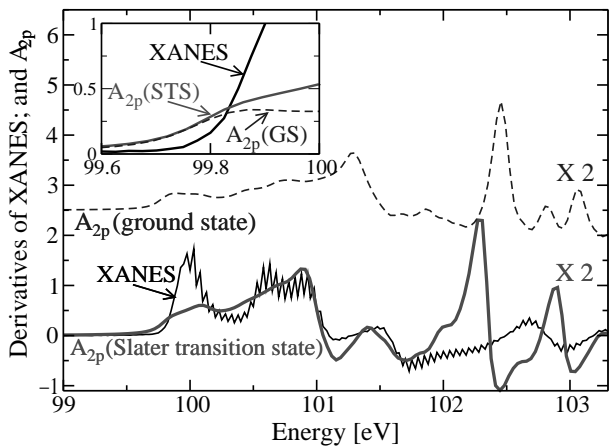}
 \caption
 { (a) Derivative of XANES for c-Si, compared to the STS and GS
derivatives of $A_{2p}$, using the LAPW method.\cite{WIEN} The inset shows
the threshold region. (The theoretical curves are matched to the
experiment to within $\pm$0.15 eV.) 20 $\times$ 20 $\times$ 20 {\bf k}
points are used. The broadening is set to 0.015 eV.\cite{fuggle} }
 \label{slaterDOS}
 \end{figure}

Figure \ref{slaterDOS} shows the measured XANES derivative combined with
two theoretical Si-$2p$ absorption derivatives, one obtained from the STS
and the other from the ground states (GS), all for c-Si. Both calculations
are performed within the all-electron, linearized-augmented-plane-waves
(LAPW) framework.\cite{WIEN} The STS calculation is set by simultaneously
reducing and increasing by one-half the electronic charge of the 2$p$ core
level and the total number of valence electrons, for a Si atom in the FCC
cell. This is imposed during the whole iterative process. In this way, the
excess valence electron gets promoted to \emph{some} conduction levels The
$s$- and $d$-DOS's of c-Si are then combined with their matrix elements.
The latter are shown to vary slowly and monotonically as
$|M_{2p,d}(\omega)|^2 \simeq 0.086773 + 0.0027173*\hbar\omega$ and
$|M_{2p,s}(\omega)|^2 \simeq 0.041455 + 0.00060321*\hbar\omega$ for
energies above the Fermi level, up to $\hbar \omega\simeq$~8 eV. This is
in contrast to the results of Buczko \emph{et al.}\cite{Buczko} where a
constant and much lower ratio $|M_{2p,d}(\omega)|^2/|M_{2p,s}(\omega)|^2$
is reported.
 \begin{figure}[b]
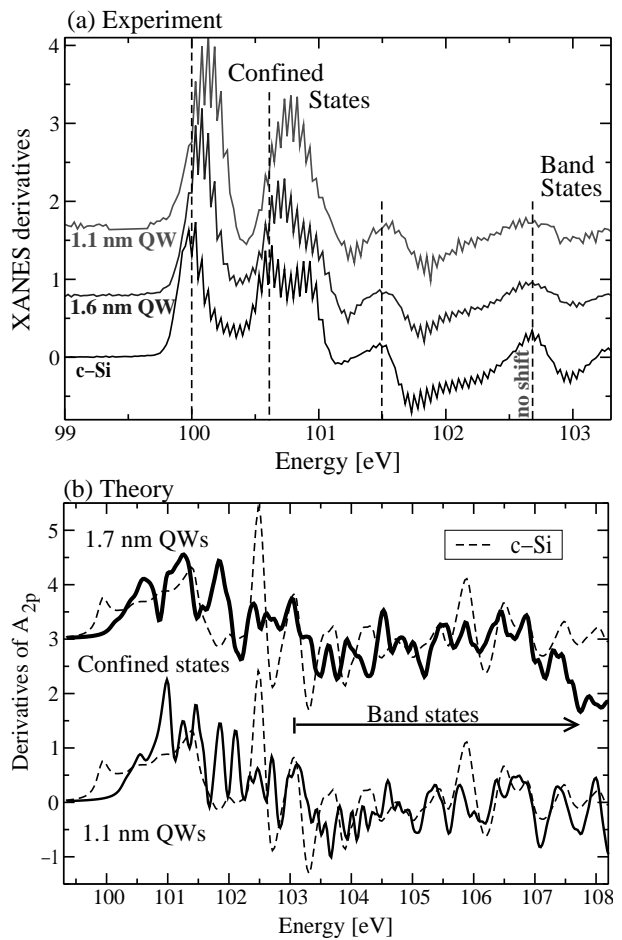

 \includegraphics*[width=8cm]{derXANESsi1.eps}
 \includegraphics*[width=8cm]{DER.eps}
 \caption
 { (a) XANES derivatives for two QW samples, and for c-Si.  (b)
Theoretical derivatives of $A_{2p}$ according to Eq.\ (\ref{A2p}), for two
QWs similar to those in (a), and for c-Si, using the PAW
method;\cite{VASP} the mean derivatives for all Si$^{0}$ atoms in the QWs
is shown. All theoretical curves are equally shifted along the XANES
threshold, for comparison with (a). 20 $\times$ 20 $\times$ 20 \textbf{k}
points are used for c-Si and 9 $\times$ 9 $\times$ 4 for both QWs (i.e.,
$\sim$1000 tetrahedra). In (b), the dotted lines correspond to c-Si.  }
 \label{xanespec}
 \end{figure}

We first observe from Fig.~\ref{slaterDOS} that above the Fermi level (at
$\sim$100 eV) up to 102 eV, the STS and the XANES are in very good
agreement.  Thus the STS is clearly an improvement on the GS calculation
as expected from equi-ensemble DFT.\cite{DreizlerGross} Second, the sharp
peak at 102.7 eV in the XANES derivative is downshifted to 102.2 eV in the
STS (as well as for the GS based calculation).  One should remark that the
XANES intensities get significantly damped for energies high above the
threshold (due to an acute angle of the photon probe tangent to the
sample), while no damping is included in the calculations. Finally, we
observe that, in spite of several shifts of the energy peaks in the GS,
the general aspect of the GS and STS theoretical curves resemble. However,
the GS is clearly expanded in the near-edge region of the spectrum (from
99.5 to 102 eV) compared to the STS, in agreement with previous
calculations (see Ref.~\onlinecite{Buczko} and reference therein);
moreover, above 102.5 eV, the GS shows several features not visible in the
STS. This comparison of the STS and GS theoretical approaches for c-Si is
essential to the analysis that follows, applied to the Si/SiO$_2$ QWs
(solely in the GS, owing to their large size as can be appreciated from
Fig.~\ref{FRM}; and hence no STS results are available).

Figure \ref{xanespec}(a) depicts the XANES derivatives of two QWs, 1.1 nm
and 1.6 nm thick, as well as for c-Si. The lowest energy edges at the
Si-conduction band minimum ($\sim$~100 eV) shifts to the blue as we go
from c-Si to the 1.1 nm Si well. This trend continues until we reach peaks
at and above $\sim$~102.7 eV. The first peaks above are seen to be
\emph{unshifted} with respect to the c-Si peaks. This is easily explained
if we assume that the Si/SiO$_2$ interfaces are associated with a CBO of
$\sim$~2.7 eV.  However, given the $\sim$~9 eV bandgap of crystalline
SiO$_2$ and the 1.1 eV bandgap of c-Si, this leads to a VBO of about 3.95
eV, as commonly assumed.\cite{williams} Thus the XANES suggests that a
large reduction in the CBO has occurred.

In order to evaluate the electronic properties of the confined states, a
calculation of $A_{2p}$ obtained from Eq.\ (\ref{A2p})  has been applied
to Si atoms inside two Si wells.  The matrix elements are deduced from
c-Si, calculated within the LAPW and the STS approaches given above, while
the $s$- and $d$-DOS's are evaluated in the GS, within the
projector-augmented-waves (PAW)  approach.\cite{VASP} Thus, in this
procedure, we assume that the matrix elements of Si$^{0}$ atoms (i.e., Si
atoms inside only the Si well)  are equivalent to their counterpart in
c-Si.  This assumption is justified by the observed slow variation with
$\hbar \omega$ of the two matrix elements $|M_{2p,d}(\omega)|^2$ and
$|M_{2p,s}(\omega)|^2$, as discussed before. Figure \ref{xanespec}(b)  
shows the mean (lorentzian) derivatives of $A_{2p}$ for Si$^{0}$ atoms in
the 1.1 nm and 1.7 nm Si/SiO$_2$ QWs, as well as for c-Si. The Si-$2p$
absorption of the QWs are presented for energies that correspond to the
$2p\to $CBM region.

We first observe from Fig.~\ref{xanespec}(b) that the absorption
intensities in the near-edge region, from 100 eV to $\sim$102.7 eV, for
both QWs, are much higher than for c-Si (with their respective energy gaps
preserved) leading to enhanced absorption properties\cite{pcarFRM} and
strong confinement effects in this region. Second, a comparison of the two
QWs together in the ``band states'' region from $\sim$103.5 eV to higher
energies gives in general good agreement between each other. However, a
complete correspondence between any of the QWs with c-Si only appears at
energies above $\sim$105.7 eV; for instance, some features appearing in
both QWs (e.g., a triplet of decreasing peaks at 105 eV)  are not present
in c-Si. Thus, the electronic properties of silicon in confined structures
are modified high above the CBO. In the region between 102.7 eV and 105.7
eV, both QWs show rather similar features, while no complete agreement
lies between any of the QWs with c-Si. This suggests a ``transition
domain'' for the band states region where the QWs together have similar
electronic properties but different from c-Si.

Thus, within the GS, two main regions are observed in the QWs, as from the
XANES:  one having strong confined states, and another region of band
states constituted of a transition domain where their electronic
properties are slightly different from c-Si and for much higher energies a
region of pure c-Si band states.  It is important to recall that the
absorption calculations presented in Fig.~\ref{xanespec}(b) for the QWs
(and c-Si) do not incorporate any approximations for the excited states,
such as the STS approach used for c-Si as shown in Fig.~\ref{slaterDOS};
informations contained in Fig.~\ref{xanespec}(b) thus remain qualitative.

In summary, the detailed electronic structures within confined crystalline
wells have been measured for the first time. The XANES data show rich
electron states in these crystalline silicon wells, in dramatic contrast
to featureless XANES from amorphous silicon wells.\cite{llb} The data show
dramatic blue shift for electron states at the band edge (i.e., the bottom
of the well). The magnitude of the shift decreases as the energy levels
increase and eventually there is no shift for states above $\sim$2.7 eV
from CBM, as one would expect from elementary quantum mechanics for a
finite-well system. On the theory side, we have built detailed atomic
models for the SiO$_2$/Si/SiO$_2$ system. Based on such realistic models,
we have developed theoretical methodologies to calculate XANES data.  
Ground states calculations for the theoretical QWs give a confined states
region much higher in energy than in experiment, with a transition domain
where electronic properties of the QWs are together alike but different
from c-Si.  Strong confinement has been confirmed both from experiment and
the present ground-state study, that provides qualitative insights on the
electronic properties, i.e., for excited states. An excited states
description applied to the quantum wells, such as the
Slater-transition-state theory applied here to the c-Si, would provide
further accurate knowledge of the confinement effects in these Si
nanostructures.

{\it Acknowledgments} -- This work is supported by grants from the NSERC
of Canada and the FCAR of the Province of Qu{\'e}bec. We are indebted to
the RQCHP for generous allocations of computer resources. We thank CANNON
ELTRAN for providing us the SOI wafers.

\end{document}